\documentclass[12pt,a4paper]{article}
\pdfoutput=1
\usepackage[format=hang]{caption}
\usepackage{epsfig,amssymb,amsmath,graphicx,subcaption,verbatim,hyperref,ulem,bm,float,tensor,tikz}
\usepackage{blkarray}
\usetikzlibrary{decorations.markings}
\newcommand{\be}{\begin{equation}}
\newcommand{\ee}{\end{equation}}

\newcommand{\news}{\setcounter{equation}{0}}
\def\bea{\begin{eqnarray}}
\def\eea{\end{eqnarray}}
\numberwithin{equation}{section}
\makeatletter
\renewcommand*\env@matrix[1][\arraystretch]{
  \edef\arraystretch{#1}
  \hskip -\arraycolsep
  \let\@ifnextchar\new@ifnextchar
  \array{*\c@MaxMatrixCols c}}
\makeatother
\usepackage{color}

\begin{document}

\title{\vskip -65pt
\vskip 60pt
{\bf {\large Entanglement Entropy of Topological Orders with Boundaries}}\\[20pt]}
\author{\bf {Chaoyi Chen$^3$, Ling-Yan Hung$^{1,2,3}$, Yingcheng Li$^3$, Yidun Wan$^{1,2,3,4,5}$}\\[20pt]
$^1$State Key Laboratory of Surface Physics\\ 
Fudan University,\\
200433 Shanghai, China\\
$^2$Collaborative Innovation Center of Advanced Microstructures,\\
210093 Nanjing, China\\
$^3$Department of Physics and Center for Field Theory and Particle Physics,\\ Fudan University\\
200433 Shanghai, China\\
$^4$Institute for Nanoelectronic devices and Quantum computing,\\ Fudan University,\\ 200433 Shanghai , China\\
$^5$Department of Physics and Institute for Quantum Science and Engineering,\\ Southern University of Science and Technology,\\ Shenzhen 518055, China
}

\date{\today}
\maketitle
\vskip 20pt

\begin{abstract}
In this paper we explore how non trivial boundary conditions could influence the entanglement entropy in a topological order in 2+1 dimensions. Specifically we consider the special class of topological orders describable by the quantum double. We will find very interesting dependence of the entanglement entropy on the boundary conditions particularly when the system is non-Abelian. Along the way, we demonstrate a streamlined procedure to compute the entanglement entropy, which is particularly efficient when dealing with systems with boundaries. We also show how this method efficiently reproduces all the known results in the presence of anyonic excitations.
\end{abstract}

\vfill

%PACS:
%\newpage

\section{Introduction}\news

Entanglement entropy has been a powerful probe to detect properties of matter.
For example, entanglement entropy can probe different topological orders.
In 2+1 dimensions, the entanglement entropy of a topologically ordered quantum system, when the system is divided into two regions $A$ and $B$, takes the following generic form
\be
S(A) = \alpha \frac{L_A}{\epsilon} -\gamma,
\ee
where $L_A$ is the length of the boundary of region $A$, $\epsilon$ the UV cutoff, and $\gamma$, the universal term
named ``topological entanglement entropy'' \cite{Levin:2006zz, Kitaev:2005dm}. It is known that when $S(A)$ is evaluated on the ground state of the system on a sphere,
\be \gamma = n \ln D, \ee 
for some region $A$ consisting of $n$ disconnected disks, and $D$ the total quantum dimension of the topological order, defined as
\be
D = \sqrt{\sum_i d_i^2},
\ee
and $d_i$ are the quantum dimensions of the anyons $a_i$ of the topological order. 

In the current paper, we shall inspect the entanglement entropy in a topologically ordered system (to be called a topological order for simplicity unless otherwise stated) with non-trivial boundary conditions.  

To avoid any confusion, we shall refer to a boundary between two regions in
a system as an entanglement boundary (EB) and a physical, gapped boundary between the system and vacuum a physical boundary (PB). 

Non-chiral topological orders can admit gapped, or alternatively topological boundary conditions. In 2+1 dimensions, each such boundary condition is related to an algebra often called the ``Lagrangian subalgebra'' \cite{Levin:2013gaa, Kapustin:2010hk, Wang:2012am, Lan:2014uaa}. It is also characterized by a Frobenius algebra \cite{Fuchs:2003yk} in the modular tensor category describing the topological order or by a Frobenius algebra \cite{Hu:2017lrs,Bullivant:2017qrv, Hu:2017faw} in the unitary fusion category describing the fundamental degrees of freedom of the topological order. These boundary conditions are also known to correspond to the physics of ``anyon condensation''\cite{Bais:2008ni, Eliens:2013epa, Kong:2013aya,  Hung:2014tba}. Each such boundary is also associated to some modular invariant \cite{Levin:2013gaa, Fuchs:2003yk, Hung:2015hfa}. 
We would like to inspect whether the entanglement entropy can detect these boundary conditions, and if so, whether the resultant values correspond to certain topological invariants. 

We will inspect this problem in the context of the twisted quantum double (TQD) models of $(2+1)d$ topological orders\cite{Hu:2012wx}, which are a Hamiltonian extension of the Dijkgraaf-Witten topological gauge theories \cite{Dijkgraaf:1989pz}. The entanglement entropy of these models have been computed before, notably in \cite{Hamma1, Flammia:2009axf} in the absence of physical boundaries. In the case of $Z_2$ toric code model, the case with PB was also briefly discussed in \cite{Zhang:2011jd}.
We will revisit the problem, and introduce a streamlined method.  It involves reducing the problem systematically to one that is independent of system size. We also clarify the construction of Schmidt decomposition by systematically choosing a convenient canonical set of basis. As we will see, the modified discussion would enable an efficient and clear inspection of the scenario when we have non-trivial boundary conditions. 

\section{TQD (Dijkgraaf-Witten) models }

Dijkgraaf-Witten topological gauge theories were formulated initially as a way of defining a path-integral of a discrete version of Chern-Simons theory. They were later adopted for defining quantum Hamiltonians, whose ground states admit exotic properties, that we now identify as the fixed-point wavefunction of topological orders. 
For a detailed discussion, we refer the readers to \cite{Kitaev:1997wr, Hu:2012wx}. 

We only collect the necessary ingredients of the TQD models that would facilitate our exposition in the following. 
We note that the model is defined on a lattice $\Gamma$. The lattice does not have to be regular but without loss of generality, we shall consider a square lattice.
There is a Hilbert space $\mathcal{H}_l$ defined at each link $l$. 

We first consider a special subset of the TQD models, where there is no twist. Such models are called the Kitaev models or quantum double (QD) models. A QD model with a finite gauge group $G$, $\dim\mathcal{H}_l=|G|$, where one can choose a basis such that each basis state can be labeled by a group element of the group $G$,
has the Hamiltonian
\be
H = - \sum_v A_v - \sum_p B_p, \qquad A_v= \frac{1}{|G|} \sum_g A_v (g), \qquad B_p = B_p(e).
\ee
The subscript $v$ denotes vertices, and $p$ denotes plaquettes.
The action of $A_v(g)$ and $B_p(g)$ is illustrated in figure \ref{fig:AB}.

\begin{figure}[h]
\centering
\begin{tikzpicture}[scale=0.6]
\begin{scope}[very thick,decoration={markings,mark=at position 0.5 with {\arrow{>}}}] 
\draw[postaction={decorate}] (2,2)--(4,2);
\draw[postaction={decorate}] (4,2)--(6,2);
\draw[postaction={decorate}] (4,2)--(4,4);
\draw[postaction={decorate}] (4,2)--(4,0);
\node at (1,2) {$A_v(g)$};
\node[above] at (3,2) {$d$};
\node[above] at (5,2) {$b$};
\node[right] at (4,1) {$a$};
\node[right] at (4,3) {$c$};
\node at (7,2) {$=$};

\draw[postaction={decorate}] (8,2)--(10,2);
\draw[postaction={decorate}] (10,2)--(12,2);
\draw[postaction={decorate}] (10,2)--(10,4);
\draw[postaction={decorate}] (10,2)--(10,0);
\node[above] at (9,2) {$dg^{-1}$};
\node[above] at (11,2) {$gb$};
\node[right] at (10,1) {$ga$};
\node[right] at (10,3) {$gc$};
\node at (13,2) {$,$};
\end{scope}
\end{tikzpicture}

\begin{tikzpicture}[scale=0.8]
\begin{scope}[very thick,decoration={markings,mark=at position 0.5 with {\arrow{>}}}] 
\draw[postaction={decorate}] (1,0)--(3,0);
\draw[postaction={decorate}] (3,0)--(3,2);
\draw[postaction={decorate}] (3,2)--(1,2);
\draw[postaction={decorate}] (1,0)--(1,2);
\node[left] at (0,1) {$B_p(g)$};
\node[right] at (3,1) {$b$};
\node[above] at (2,2) {$c$};
\node[left] at (1,1) {$d$};
\node[below] at (2,0) {$a$};
\node at (4,1) {$=$};

\draw[postaction={decorate}] (7,0)--(9,0);
\draw[postaction={decorate}] (9,0)--(9,2);
\draw[postaction={decorate}] (9,2)--(7,2);
\draw[postaction={decorate}] (7,0)--(7,2);
\node at (5.5,1) {$\delta_{g,abcd^{-1}}$};
\node[right] at (9,1) {$b$};
\node[above] at (8,2) {$c$};
\node[left] at (7,1) {$d$};
\node[below] at (8,0) {$a$};
\node at (10,1) {$.$};
\end{scope}
\end{tikzpicture}
\caption{An illustration of the action of $A_v(g)$ and $B_p(g)$.}
\label{fig:AB}
\end{figure}

Since all the operators $A_v$ and $B_p$ commute, the ground state can be generated by
\be \label{GS}
|\psi\rangle =\prod_v  A_v | \Omega\rangle,
\ee
where $|\Omega\rangle$ is some appropriate reference state satisfying $B_p |\Omega\rangle = | \Omega\rangle$. When the (closed) two dimensional space on which the state lives is a sphere, the reference state $|\Omega\rangle$ is simply given by all links taking the state corresponding to the identity element $e$ of the group $G$, namely
\be \label{GS1}
|\Omega\rangle=\bigotimes\limits_{l} |e\rangle_l,
\ee
where $l$ runs over all links of the lattice $\Gamma$. This state $|\Omega\rangle$ is of course a direct product state. When the $2$-dimensional space has non contractible cycles, there would be a ground state degeneracy, and a basis of the degenerate ground states can be constructed by taking a set of reference states with closed ribbon operators  acting on non-contractible cycle in the trivial reference state $|\Omega\rangle$. 

The analysis in the following for individual such states are all the same. We will discuss general linear combinations of these basis in later sections when we encounter the geometry of a cylinder. 

\subsection{Gapped boundaries}
The PB conditions of the QD models have been discussed in \cite{kitaev_kong, 2011CMaPh.306..663B} and subsequently generalized to the TQD models \cite{Bullivant:2017qrv,Hu:2017w} , to the Levin-Wen models \cite{Hu:2017lrs,Hu:2017faw}, and to higher dimensions \cite{Yoshida:2017xqa}. 

We will illustrate our methods mainly using the QD model. 
Each PB is characterized by a subgroup $K \subset G$ \cite{2011CMaPh.306..663B}. In fact, it is shown in \cite{Bullivant:2017qrv} that even for a TQD model with a gauge group $G$, a PB condition is fully characterized by a subgroup $K\subseteq G$.

The PB Hamiltonian is given by 
\be
H_B = - \sum_{v_B}A_{v_B} - \sum_{l_B} B_{l_B}
\ee
where $A_{v_B}$ acts on the links connected to the vertex located at a PB given by $A_{v_B} = \frac{1}{|K|}\sum_{k\in K} A_{v_B}(k)$, and $B_{l_B}$ is a projector on the PB links to the subgroup $K$.

A ground state is generated in a similar manner as in (\ref{GS}), except that for vertices on the PB we replace a generic $A_{v}$ by $A_{v_B}$.

\section{Revisiting the entanglement entropy of the QD models}

In this section, we revisit the problem of computing the entanglement entropy in the QD models.  We shall lay out a procedure that is improved compared with the discussion in \cite{Hamma1, Flammia:2009axf, Hung:2015fla}. 
The procedure would allow one to obtain the entanglement entropy in the presence of PBs in a systematic and clear manner.

Now for simplicity, we consider again the case of an entangling region $R$ taking the shape of a disk on the sphere.

On the sphere, any topological order has unit ground state degeneracy. The ground state on the sphere is generated as described above, in equation (\ref{GS}).  It is known that the operators $A_v$ acting on $v$ away from the  EB between the region $R$ and its complement $\bar R$ do not contribute to generating entanglement between the regions.
It is also known that the entanglement arises from operators $A_{v_b}$ that act on the vertices $v_b$ along the EB  and hence affect both region $R$ and  $\bar R$ at the same time. Hence, one can simply label some $i$-th EB configuration by the set of vertex operators $\{A_{v_b}( {g_{i}}_b )\}$. 

Consequently, one can just focus on different EB configurations $\{A_{v_b}( {g_{i}}_b )\}$ in each term in the ground-state wavefunction. As usual, there is a physical ambiguity over the definition of entanglement entropy on a lattice gauge theory. But here, we have taken the viewpoint explained for example in \cite{Buividovich:2008gq, Donnelly:2011hn,Ghosh:2015iwa, Aoki:2015bsa,Soni:2015yga}, and work with the extended Hilbert space, which should agree with the electric center in terms of the choice of operator algebra in the original gauge theory \cite{Casini:2013rba}.

The game is to obtain a Schimidt decomposition to recover the reduced density matrix. Given that only $A_{v_b}$ are responsible for the entanglement between $R$  and $\bar R$, a naive Schimidt decomposition is obtained as
\be \label{decompo1}
|\Psi\rangle = \frac{1}{|G|^L} \sum_{i=1}^{|G|^L} \left[ |R_{i}\rangle \otimes  |\bar R_{i}\rangle \right], 
\ee 
where
\be \label{decompo2}
|R_{i} \rangle \otimes |\bar R_{i} \rangle =   \prod_{b=1}^L A_{v_b}({g_i}_b) \left[ \sum_{\{ A_{v_r}\}} \prod_{v_r\in R} A_{v_r} |0\rangle_R \otimes  \sum_{\{A_{v_r'}\}}\prod_{v_{r'}\in \bar R} A_{v_{r'}} |0\rangle_{\bar R}\right],
\ee
for some $i$-th EB configuration $\{A_{v_b}( {g_{i}}_b )\}$, where $L$ is the length, i.e., the number of links, along the EB. We note that the rhs above is indeed a direct product state in $R$ and $\bar R$, agreeing with the expression on the lhs,  since $\prod_{b=1}A_{v_b}({g_i}_b)$ is a tensor product of operators acting on $R$ and $\bar R$ respectively. 
Had each EB configuration $\prod_b A_{v_b}(g_b)$ always led to a pair of orthogonal internal states $|R_i\rangle $ and  $|\bar R_i\rangle$, the Schidmit decomposition would have been completed. This is however not the case. Indeed, two different sets $\{A_{v_b}({g_{i,j}}_b)\}_{i, j}$  generating states $|R_{i,j}\rangle$ and similarly $| \bar R_{i,j}\rangle $ for different pairs of $i,j$  are generically not linearly independent.  That is, our naive labelling of the internal states based on the EB configuration of $A_{v_b}$ does not in general lead to $|G|^L$ orthogonal $|R_i\rangle$ and $|\bar R_i\rangle$ separately. 
There is a complication, namely
that {many of the $|R_i\rangle$ ($|\bar R_i\rangle$) for different $i$ correspond to the same state.} This is because the global action of the internal $A_{v\in R \,\,(\bar R)}$ in each connected component of $R \,\,(\bar R)$ (excluding $A_{v_b}$) could mimic the effect of the global action of $A_{v_b}$ within links in the component. (See the detailed exposition in section \ref{sec:interp} on the bath of ribbons generated by $A_v$. )

As we are going to see, more generically when the region $R$ or $\bar R$ has topologies more non-trivial than a disk, or when the system is placed on surfaces beyond a sphere, the ground state is still generated by in an analogous manner as in \ref{GS} with the reference state $|\Omega\rangle$ potentially dressed by ribbon operators (the discussion of these dressed states are postponed until section \ref{sec:genericribbon}). The paremetrization adopted in \ref{decompo1} and \ref{decompo2} can still work quite generally. Generically, we would be met with a new complication.  
{That is, whenever we transform a state by $A_{v}(g)$ at all $v$ within region $R$, including the entanglement boundary ${v_b}$ at the same time, we would keep $|R_i\rangle$ invariant, while taking $|\bar R_i \rangle$ to another state $|\bar R_j \rangle$.  }The EB configuration after such a transformation has been shifted $ \{A_{v_b}(g_b)\} \to \{A_{v_b}(g_b g)\} $ however.  This implies that some of the $|R_{i,j}\rangle $ with $i \neq j$ may in fact be the same but they do connect to different $|\bar R_{i,j}\rangle$, leading to entanglement pattern in the wavefunction of the form $|R_i\rangle \otimes (|\bar R_i \rangle + |\bar R_j \rangle +\cdots)$.

In the classic literature on the subject\cite{Hamma1}, further analysis is based on defining a huge group $\mathcal{G}$ corresponding to the action of $A_v(g)$ on the entire lattice, and attempting to obtain the quotient group $\mathcal{G}_{R,\bar R}$, where we quotient by the action of the corresponding groups $\mathcal{G}_R$ and $\mathcal{G}_{\bar R}$, where $\mathcal{G}_X$ includes action of $A_v$ on vertices $v$ within $X$, excluding the EB. This is of course Mathematically correct but the analysis on complicated situations where there are PBs can be confusing at times.

\subsection{A modified analysis}
The improvement proposed in the current paper is to consider explicitly a division of  the collection of $|G|^L$ EB configurations  $\prod_b A_{v_b}(g_b)$ into distinct sets. 
The choice of such a division is such that each group would contribute to identical blocks  in the reduced density matrix. i.e. The complications that arise due to global applications of $A_v(g)$ within $R$ or $\bar R$ described above could generate off-diagonal elements only within each block. 

%Consider a boundary of $N$ disconnected components.  For any given configuration of $\prod_i^N \prod_{b_i} A_{v_b}(g_{b_i})$ on the boundary, where $b_i$ distinguishes the different disconnected components of the boundary, 
%we collect all configurations related to the reference configuration by $\prod_i^N \prod_{b_i}A_{v_b}(g_{b_i} g_i)$ into a set. i.e. we collect boundary configurations related to each other by at most a global shift by some element $g_i$ on each individual boundary.
%There are $|G|^N$ members in a set, and there are $|G|^{L-N}$ separate sets that would not interfere with each other as we mod out actions of global transformations in various regions. Individual member within a set can then be labelled by an $N$-tuple: $(g_1 , \cdots g_N)$.  We call each of this set a $G^N$- orbit. 
%The reduced density matrix would be reduced to a block diagonal form with $|G|^{L-N}$ block, each block being identical. Now we analyze the problem of Schimidt decomposition within this block.

%This division is based on the observation that global application of $A_v$ either within region $R$ or its complement would only lead to $A_{v_b}(g_b * g)$ over the entire boundary. 
%Therefore, identical internal states may get  connected to different sets of $A_{v_b}(g_b)$ related to each other by $A_{v_b}(g_b * g)$.  If the boundary of region $R$ contains multiple disconnected components, then it is possible that $A_{v_b}$ corresponding to different components get shifted by different group elements $g$.
%

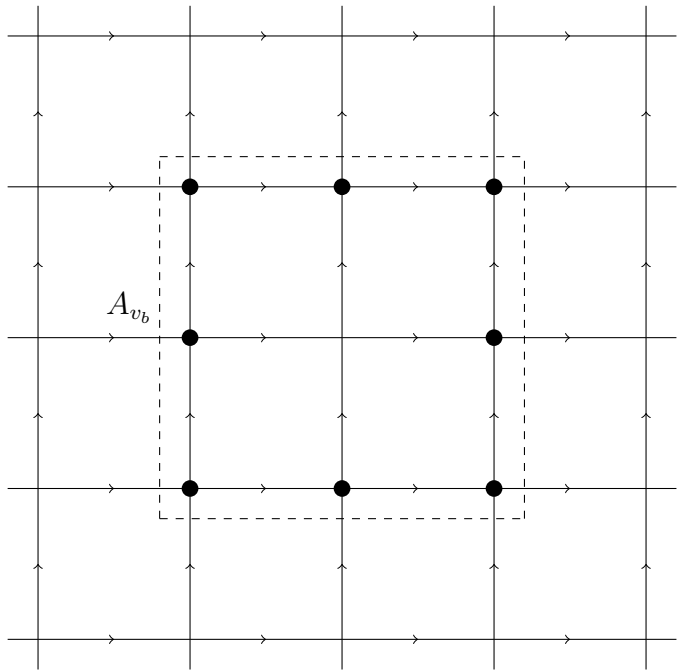
\begin{figure}[h]
\centering
\begin{tikzpicture}[scale=2]
\begin{scope}[decoration={markings,mark=at position 0.5 with {\arrow{>}}}]
\foreach \i in {1,...,4} {
  \draw[postaction={decorate}] (\i,5) -- (\i +1,5);
  \draw[postaction={decorate}] (5,\i) -- (5,\i+1);
  \foreach \j in {1,...,4} {
    \draw[postaction={decorate}] (\i, \j) -- (\i + 1, \j);
    \draw[postaction={decorate}] (\i, \j) -- (\i, \j + 1);
  }
}
\end{scope}
\foreach \i in {1,...,5} {
  \draw (0.8, \i) -- (1,\i);
  \draw (\i,0.8) -- (\i,1);
  \draw (5, \i) -- (5.2,\i);
  \draw (\i,5) -- (\i,5.2);
}
\draw[dashed] (1.8,1.8) -- (1.8,4.2);
\draw[dashed] (1.8,1.8) -- (4.2,1.8);
\draw[dashed] (4.2,1.8) -- (4.2,4.2);
\draw[dashed] (1.8,4.2) -- (4.2,4.2);

\fill (2,2) circle[radius=0.055];
\fill (2,3) circle[radius=0.055];
\fill (2,4) circle[radius=0.055];
\fill (3,2) circle[radius=0.055];
\fill (4,2) circle[radius=0.055];
\fill (4,3) circle[radius=0.055];
\fill (4,4) circle[radius=0.055];
\fill (3,4) circle[radius=0.055];

\node at (1.6,3.2) {$A_{v_b}$};
\end{tikzpicture}
\caption{Region $R$ is enclosed by the dashed line. It has the topology of a disk. }
\label{fig:disk1}
\end{figure}

These independent blocks are obtained as follows. Let us begin with the simplest scenario, where the model is defined on a 2-sphere. Region $R$ is a disk with circumference  $L$ on the sphere. That is, the EB between $R$ and $\bar R$ consists of $L$ links. This is illustrated in figure \ref{fig:disk1}.

Then there are $|G|^{L-1}$ sets. Each set is obtained in this way: take a configuration along the EB as a reference, which is a result of acting the vertex operators $\Pi_{i\in EB} A_{v_i}(g_i)$, then multiply each $g_i$ with one and the same group element $g\in G$ to reach another configuration in the set, and repeat this for all element of $G$. This is the $G^1$-orbit of configurations along the EB and contains precisely $|G|$ configurations referred to above. Each configuration in a $G$-orbit is clearly labeled by certain group element  $g\in G$. 
According to the discussion above therefore, we write
\be
|\psi\rangle_{\textrm{1 block}} = \frac{1}{|G|}\sum_i^{|G|}  | R(g_i ) \rangle \otimes |\bar R(g_i)\rangle,
\ee
where $g_i$ labels the global shift along $v_b$ over a reference representative EB configuration in a $G$-orbit. 

Now in this case, the entanglement boundary is contractible both within $R$, and in $\bar R$. We note that each extra global action of $g_i$ at $A_{v_b}$ is to create a pair of $g_i$ shifts that form a closed loop in both $R$ and $\bar R$ simultaneously, and they are contractible since the EB is contractible. 
This immediately suggests that $|R \,\,(\bar R)\,\, (g_i)\rangle = |R\,\, (\bar R)\,\, (g_j) \rangle $ for all $i,j$, due to the separate action of $\prod_{v_r\in R \,\,(\bar R)} A_{v_r}$ within the regions.

We thus conclude that the wavefunction within this $G$-orbit can be simplified to
\be
|\psi\rangle_{\textrm{1 block}} =  | R(g_1\rangle) \otimes  |\bar R(g_1 \rangle),
\ee
which is a direct product state within the block. Therefore each block contributes to a 1 dimensional projector to the reduced density matrix.

The entanglement entropy thus reads 
\be
S(R) = \ln |G|^{L-1} = \ln |G|^L - \ln |G|,
\ee
which is just the log of the number of blocks and recovers the well-known result.

Now, consider an EB of $N$ disconnected components, each component contractible in both $R$ and $\bar R$.  For any given configuration of $\prod_i^N \prod_{b_i} A_{v_b}(g_{b_i})$ on the EB, where $b_i$ distinguishes the different disconnected components of the EB, 
we collect all configurations related to the reference configuration by $\prod_i^N \prod_{b_i}A_{v_b}(g_{b_i} g_i)$ into a set. i.e. we collect EB configurations related to each other by at most a global shift by some element $g_i$ on each individual EB.
There are thus $|G|^N$ members in a set, which is a disjoint union of $N$ $G$-orbits. There are $|G|^{L-N}$ separate $G^N$-orbits that would not interfere with each other as we mod out actions of global transformations in various regions. The reduced density matrix would be reduced to a block diagonal form with $|G|^{L-N}$ blocks, each being identical. 

Individual member within a $G^N$-orbit can thus be labelled by an $N$-tuple: $(g_1 , \cdots g_N)$.  
Using the same reasoning, for $N$ disconnected  EBs,  we have $|G|^{L-N}$ blocks, while each block is again equivalent to a single direct product state.  We thus obtain 
\be
S(R_N) = \ln |G|^{L-N} = \ln |G|^L - N \ln |G|
\ee
which is indeed the well known result for the entanglement entropy.

This method can also be readily applied to compute entanglement entropy in the presence of anyon excitations. We will for illustrative purpose demonstrate these applications in the appendix.

To summarize, this method systematically reduces a problem of treating a huge density matrix that scales as $|G|^L$ for an entanglement boundary of size $L$, to one whose dimension only scales at most with $|G|^N$, where $N$ is the number of disconnected components of the entanglement boundary, allowing a clear analysis even in complicated situations. 

We will now apply this set of methods to the case where there are PBs and where the analysis can get substantially more complicated and the advantage of the method more pronounced.

\section{Entanglement entropy for different PBs}
In this section, we would like to apply our trick of entanglement entropy computation to the case where there are PBs. 
The choice of the PB condition would modify the final results of the entanglement entropy. As to be seen, the topological entanglement entropy has a subtle dependence on the PBs.

\subsection{Case I: Region $R$ is a disk away from the PB of a disk}
\begin{figure}[h]
\centering
\includegraphics[width=12cm]{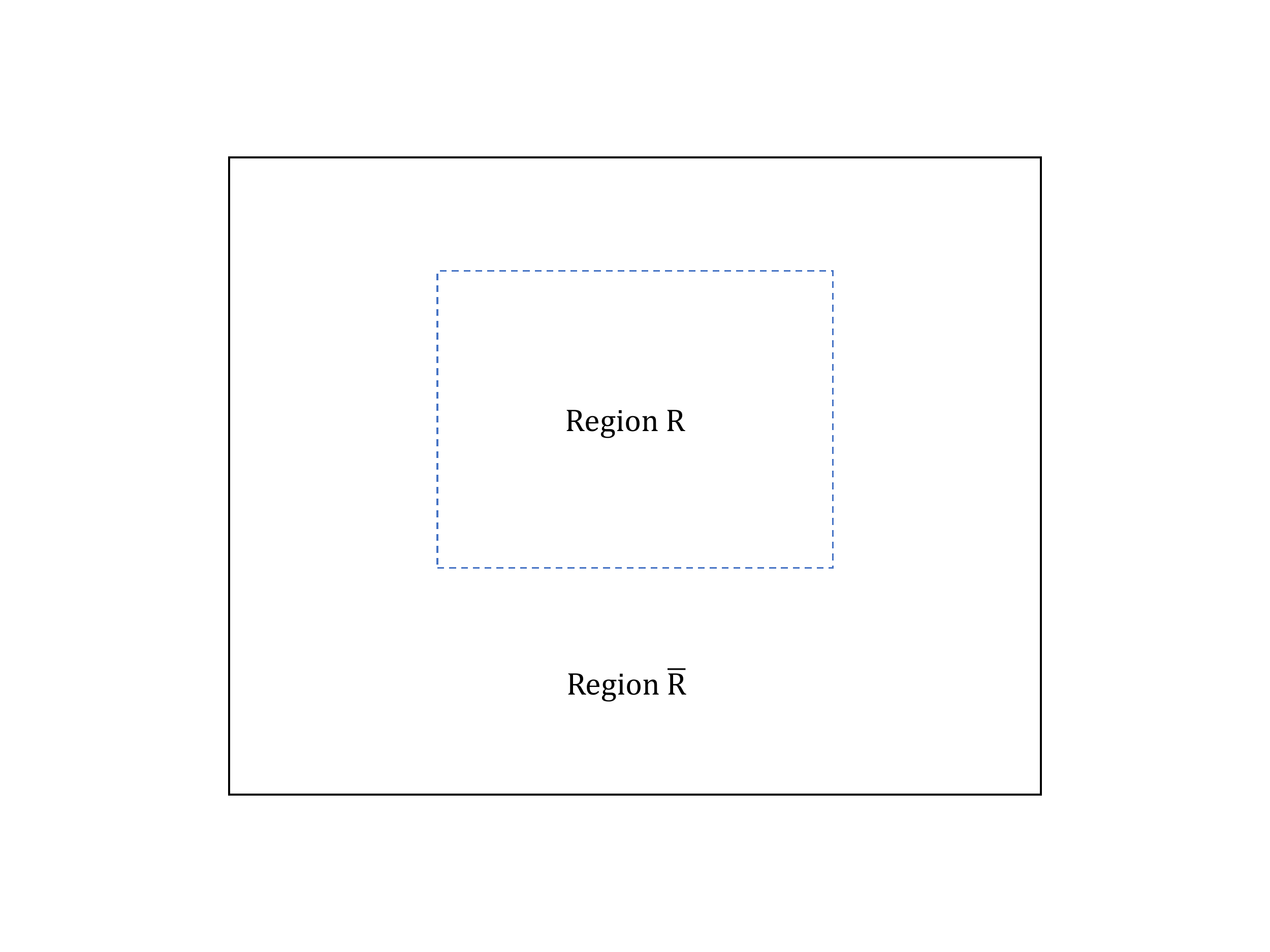}
\caption{a disk away from the PB of a disk}
\label{fig:diskdisk}
\end{figure}

Consider a disk whose PB is characterized by a subgroup $K\subseteq G$. Then consider a region $R$ that is also a disk away from the PB.  This is illustrated in figure \ref{fig:diskdisk} for R consisting of a single connected component.

In this case, we have one EB, so we divide the EB configurations into $|G|^{L-1}$ $G$-orbits, each $G$-orbit has $|G|$ members labeled by $(g)$.
Now, for precisely the same reason as in the previous analysis of a region with a disk topology, all $|R (g_i)\rangle = |R (g_j)\rangle$ for all $i,j$.  
This immediately implies that within each $G$-orbit we have a direct product state. 

The entanglement entropy is thus still given by counting the number of individual blocks, and recovers
the result
\be
S(R) = \ln |G|^{L-N}.
\ee
We have generalized the result directly to the case where $R$ contains $N$ disconnected disks. The result is insensitive to the presence of the non-trivial gapped boundaries.

We note however, that a new ingredient has crept in in the current situation. Although it did not change the entanglement spectrum, it would make a difference in later analysis. The new ingredient is that the EB may not be contractible in $\bar R$. Therefore, not all $|\bar R (g_i)\rangle$ are the same. It is interesting to check which of the $|\bar R (g_i) \rangle$
 are in fact orthogonal. We note that a global action of $A_{v\in \bar R} (k)$, $k \in K $ can generate a closed loop of $k$ shifts along the links connecting to the EB. This means that 
\be |\bar R (g_i)\rangle = |\bar R (g_i k) \rangle \ee 
for all $k \in K$. i.e. All $|\bar R(g_i)\rangle$ for $g_i$ belonging to the same left coset of $K$ corresponds to the same state. 
For completeness, we thus have 
\be
|\psi\rangle_{\textrm{1 block}} = |R (g_i) \rangle  \otimes  \frac{|K|}{|G|} (\sum^{|G|/|K|}_i  |\bar R (c_i ) \rangle ).
\ee
where $c_i$ is a representative of a left coset.  There are $|G|/|K|$  left cosets.

\subsection{Case II: Region $R$ touching the PB of a disk}

\begin{figure}[h]
\centering
\includegraphics[width=12cm]{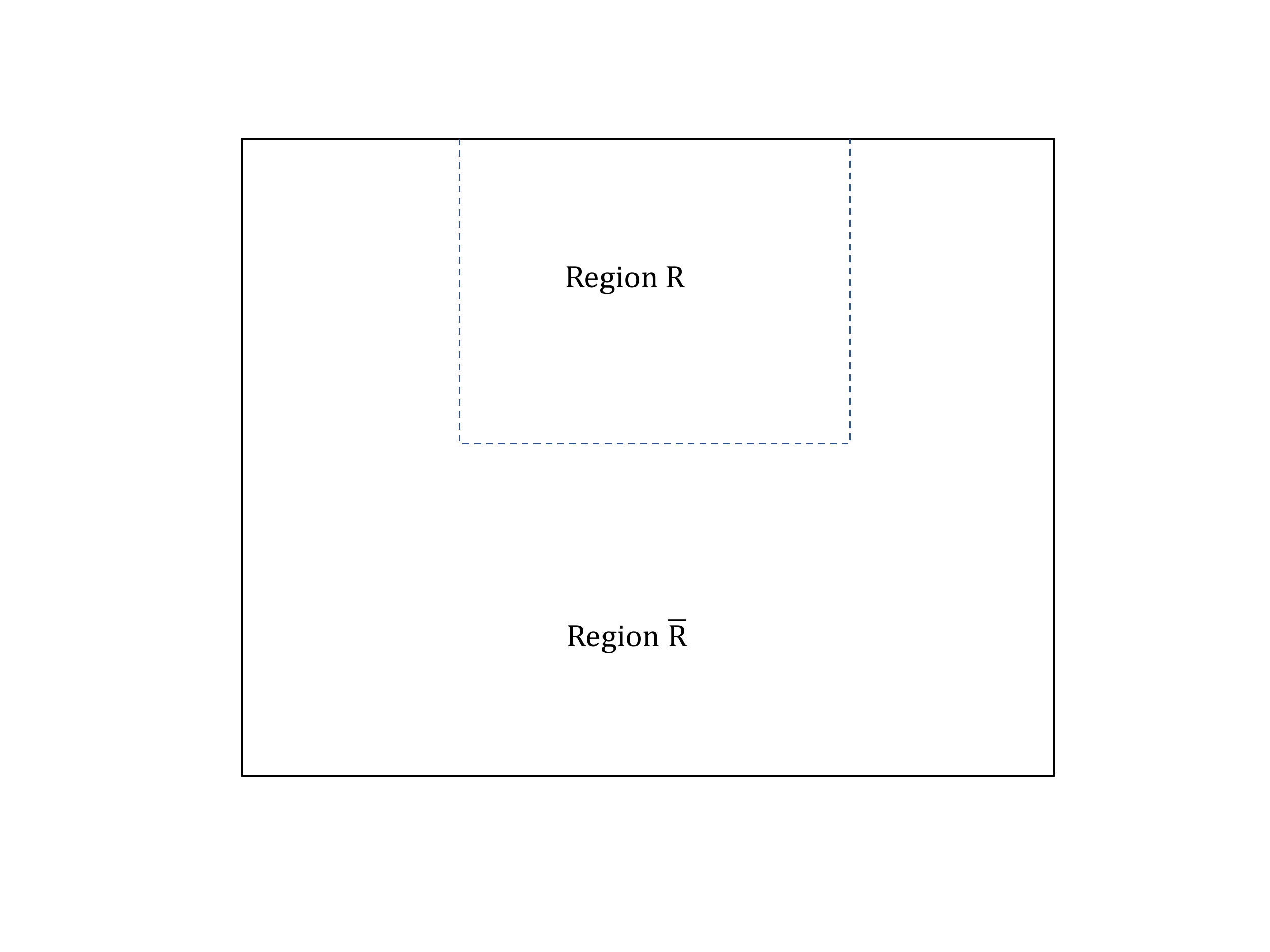}
\caption{Region R touching the PB of a disk}
\label{fig:touchdisk}
\end{figure}

Here we continue to keep the state on a disk with a PB characterized by $K \in G$. The region $R$ however, touches the PB. The EB is thus a line that begins and ends at the physical boundary.  This is illustrated in figure \ref{fig:touchdisk}.

In this case, if the EB contains $L$ vertices, then two of them sits at the PB, while the rest are located in the bulk. Therefore, the total number of configurations of possible $A_v$ sitting at the EB is given by $|G|^{L-2} |K|^2$. 
Naively we would like to divide these configurations into $G$-orbits. Nevertheless, we are not able to do so here because  $A_{v_B}(g)$ located at the PB are restricted to $g \in K$.
Instead, we can divide these configurations into $K$ orbits, with $|K|$ members in the orbit, each labeled by $(k)$. There are thus $|G|^{L-2} |K|$ distinct orbits.

In this case, the EB is not contractible in either $R$ or $\bar R$. 

From the analysis of the previous section, we note however that not all the internal $|R(g_i)\rangle$ or $|\bar R (g_i)\rangle$ are independent. They again satisfy 
\be \label{eq:identification} | R  \, \,(\bar R) \,\, (k_i)\rangle = |R \,\, (\bar R) \,\,(k_i k) \rangle \ee 
for all $k \in K$, and thus all $ | R \,\, (\bar R)\,\, (k_i)\rangle = |R\,\, (\bar R)\,\, (k_j)\rangle$ for all $i,j$.

One therefore concludes that the orbit is contributing to one direct product state.  
\be
|\psi\rangle_{\textrm{1 block}} =  ( |R (k_1)  \rangle  \otimes   |\bar R (k_1 ) \rangle ).
\ee

The entanglement entropy is then given by
\be
S(R) = \ln (|G|^{L-2} |K|)  = \ln (|K|^2 |G|^{L-2}) - \ln |K|,
\ee 
where the first term is grouped together and taken as the area term, and the second term a topological term resulting from a change in the global constraints. 
We see the first indication that the topological entanglement entropy is indeed sensitive to the physical boundary conditions.

\subsection{Case III: Region $R$ being a vertical slit on a cylinder with two PBs}

\begin{figure}[h]
\centering
\includegraphics[width=12cm]{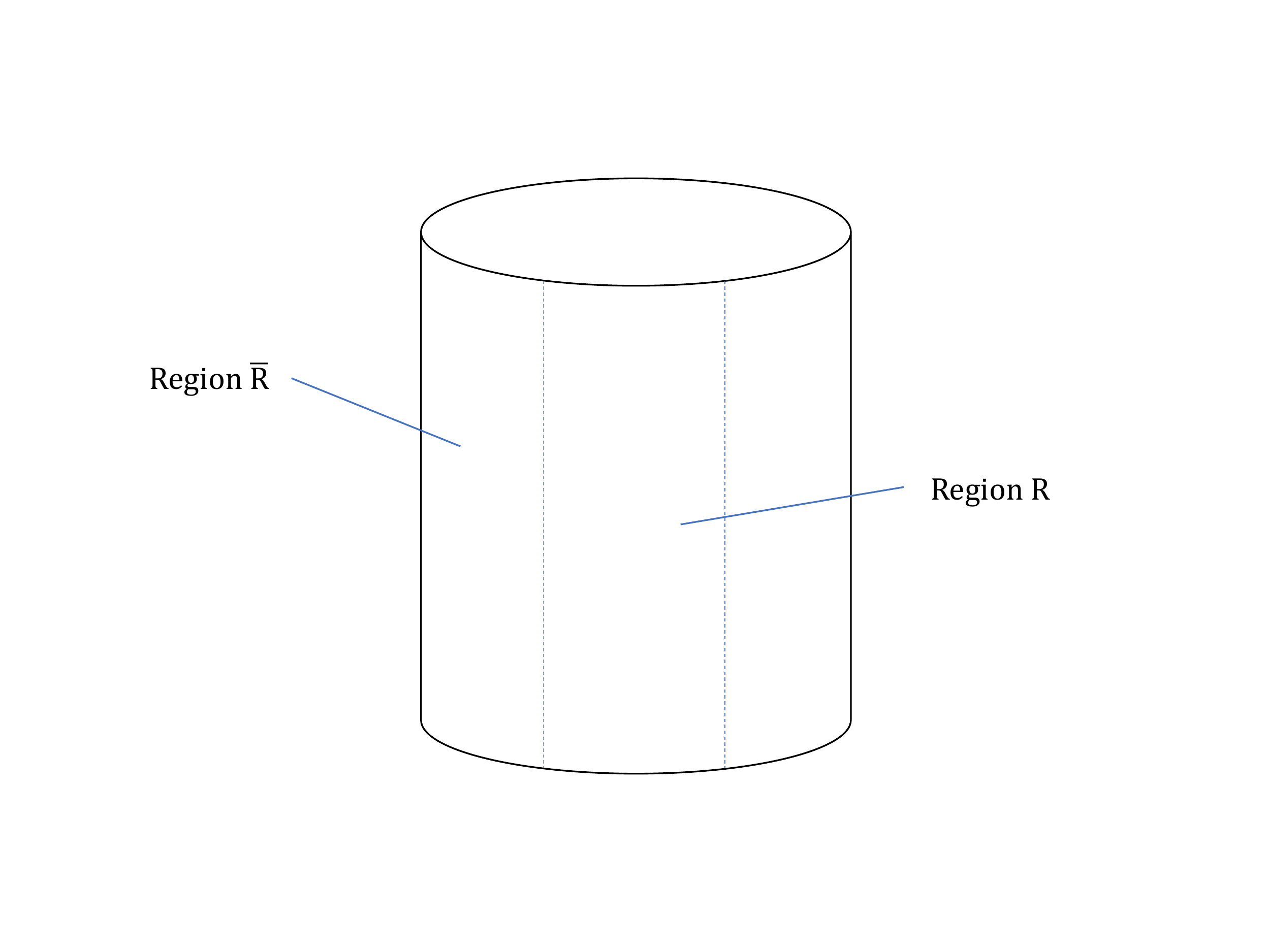}
\caption{A vertical slit on a cylinder}
\label{fig:cylinder1}
\end{figure}

Now consider a state on a cylinder with two PBs characterized by two subgroups $K_{1}$ and $K_2$ of $G$ respectively (see Fig. \ref{fig:cylinder1}). 

Now on a cylinder with a non-contractible cycle, generically there would be degenerate ground states and the entanglement entropy would depend on the precise linear combination of ground states.

It is known that one can construct a set of basis states in the degenerate ground-state subspace using ribbon operators. This is obtained by wrapping ribbon operators around non-contractible cycles. In the case where there are PBs, one can also construct basis by attaching ribbon operators that stretch between the upper and lower PBs. Not all ribbon operators can end at the PB without leading to boundary excitations, except those corresponding to condensed anyons at the PB.

We first take the simplest basis state, corresponding to no ribbon, in which the state is still describable by equation (\ref{GS}).  We will postpone the discussion of generic ground states to section \ref{sec:genericribbon}.

The first scenario we consider is a region $R$ that corresponds to a strip that connects the upper PB $K_1$ and the lower PB $K_2$. 
In this case, there are two disconnected EBs too. There are altogether $|G|^{L-4} |K_1|^2 |K_2|^2$ different EB configurations. 
For similar reasons as in the previous subsection, we are not able to divide these EBs into $G^2$ orbits because of the restriction of the physical boundary vertex. 
We are however allowed to divide the EB into $K^2$-orbit, where $K$ is the intersection of $K_1$ and $K_2$, which is itself a subgroup of $G$. Each member in the orbit is now labeled by $(k_1,k_2)$.
In this case, each connected component of the EB is not contractible, either in $R$ or $\bar R$. 

As in the previous examples, we now systematically proceed in two steps. First we deterimine which of the $|R \,\,(\bar R)\,\,(k_1, k_2)\rangle$ are identified. Then we inspect global actions in each connected component of $R$ or $\bar R$ to look for potential off-diagonal terms in the Schimdt decomposition. 

From similar analysis of non-contractible EB in the previous subsection, we conclude that
\be
| R\,\, (\bar R) \,\,  (k_1,k_2)\rangle  = | R\,\, (\bar R) \,\,  (k_1 k ,k_2 k)\rangle, 
\ee
where $K $ are group elements shared by $K_1$ and $K_2$. 

Then we look for global actions allowed within $R$ or $\bar R$ that keep the respective region invariant. Again that is restricted to elements $k$ in $K$, which also happens to take $ | R\,\, (\bar R) \,\,  (k_1,k_2)\rangle  \to  | R \,\,(\bar R) \,\, (k_1 k ,k_2 k) \rangle $.
As a result, following our procedure in the previous examples, we conclude that each $K^2$-orbit breaks up into $|K|$ entangled states. The entanglement entropy takes the form
\be
S(R) = \ln\left( \frac{|G|^{L-4}|K_1|^2|K_2|^2}{|K|^2}  |K|\right) = \ln (|G|^{L-4}|K_1||K_2| )- \ln |K|.
\ee

For region $R$ made up of $N$ such strips connecting the top PB to the bottom PB, there are $2N$ disconnected EBs. 
Repeating exactly the same analysis, we find that the entanglement entropy would take the general form
\be
S(R) = \ln (|G|^{L-4N}|K_1|^{2N}|K_2|^{2N} )-  N \ln |K|.
\ee

%[[What happens to other basis states. Electric basis. other basis.....]]

\subsection{Case IV: Region $R$ being a horizontal strip wrapping the cylinder}\label{sec:ringshape}
\begin{figure}[h]
\centering
\includegraphics[width=12cm]{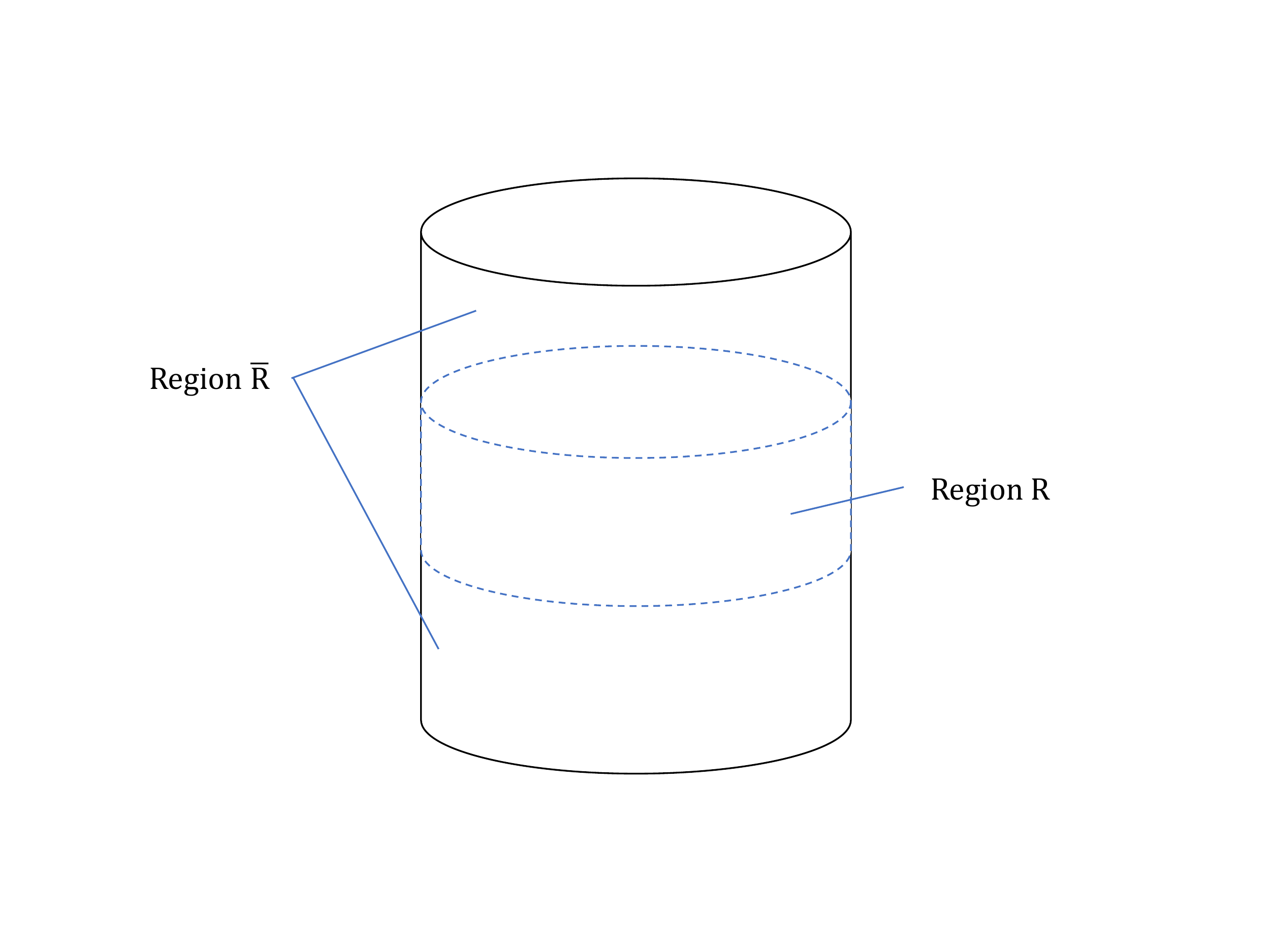}
\caption{A horizontal strip wrapping the cylinder}
\label{fig:cylinder2}
\end{figure}

Now we consider another interesting case. The state is still defined on a cylinder whose PBs are characterized by groups $K_1$ and $K_2$ respectively.

However the region $R$ is now taken as a strip wrapping the non-contractible cycle of the cylinder, separated from both PBs. 

Again, we will begin with an analysis based on the state (\ref{GS}). It turns out that this case is one that is the most interesting we have encountered so far.

The EB is made up of two disconnected components each non-contractible in either $R$ and $\bar R$. 
 As before, we first divide all the EB configurations into $G^2$-orbits with each member labelled by $(g_1, g_2)$, where $g_1$ corresponds to the upper EB and $g_2$ corresponds to the lower EB. The total number of EB configurations is $|G|^L$, where $L$ is the number of vertices on the EB, so there are $|G|^{L-2}$ $G^2$-orbits and each $G^2$-orbit has $|G|^2$ members.

Again, our first step is to determine which of the $|R\,\, (\bar R) \,\, (g_1,g_2) \rangle$ are dependent. From action within $R$ one can generate a pair of $g$ ribbons with $A_v(g)$ near the upper and lower entanglement boundary simultaneously. Similarly $A_{v}$ in region $\bar R$ can generate a ribbon $k\in K_1$ in the upper entanglement boundary and $k\in K_2$ in the lower entanglement boundary.  Therefore we have
\be \label{eq:iden}
| R (g_1, g_2)\rangle = | R (g_1g, g_2 g) \rangle, \qquad |\bar R (g_1,g_2)\rangle = |\bar R (g_1 k_1, g_2 k_2)\rangle.
\ee

Then we would like to analyze how these independent basis states of $R$ and $\bar R$ are entangled.
As in the previous analysis, global actions in $R$ together with the EB vertices preserve $R$ but shift $\bar R$. It takes the EB configuration from $(g_1, g_2) \to (g_1 g, g_2 g)$, thus taking the states $|\bar R (g_1, g_2)\rangle \to |\bar R (g_1 g, g_2 g)\rangle $.  This means that $| R( g_1,g_2)\rangle$ is paired with both $|\bar R(g_1,g_2)\rangle$ and also $|\bar R (g_1 g,g_2 g)\rangle$. These global actions form a group $G$. 
Similarly, a global action in $\bar R$ and EB preserves $\bar R$ while taking  the EB configuration $(g_1, g_2) \to (g_1 k_1, g_2 k_2)$, and thus taking  $|R (g_1, g_2)\rangle \to |R (g_1 k_1, g_2 k_2)\rangle $. These global actions form a group $K_1\otimes K_2$.

To recover a Schimdt decomposition and subsequently the reduced density matrix, we need to count the number of times each independent $| R (g_1,g_2)\rangle$ gets paired with an independent $|\bar R (g'_1, g'_2)\rangle$ after taking into account the redundancy (\ref{eq:iden}). 
To that end, we further divide these $|G|^2$ EB configurations $(g_1,g_2)$ into sets of $|G|$.  
i.e. The members in each $G^2$-orbit is further divided into $|G|$ different $G$-orbits. Members in the $G$-orbit can be related by $(g_1,g_2) = (g'_1 g, g'_2 g)$.  The EB configurations $(g_1, g_2)$ and $(1, g_2g_1^{-1})$ are in the same $G$-orbit. Therefore we can take $(1, r)$ where $r$ runs through the group $G$ as the representative of a $G$-orbit in each $G^2$-orbit. Other members of the $G$-orbit containing the representative $(1, r)$ can be parametrized by $(g, rg)$.

From (\ref{eq:iden}), it implies that all members in a given $G$ orbit are attached to the same $|R (g_1,g_2)\rangle$ state. 

Next, we would like to analyze the effect of those global actions $K_1 \otimes K_2$ that preserve $\bar R$ on the EB configurations. Under these actions, members in each $G$-orbit are allocated to the other $G$-orbits. 
For simplicity,  consider the $G$-orbit represented by $(1, 1)$. The members in that $G$-orbit are denoted by $(g,g)$. For a specific choice of $k_1$ and $k_2$, $(g, g)$ is mapped to $(gk_1, gk_2)$.  This is in the  $G$-orbit represented by $(1, gk_2k_1^{-1}g^{-1})$.  
ie. There exists a $\tilde g$ such that
\be \label{eq:conjugacy_class}
(g k_1 \tilde g, g k_2 \tilde g) = (1, gk_2k_1^{-1}g^{-1}).
\ee
That is to say, under the action of $k_1$ and $k_2$, members in the $G$-orbit $(1, 1)$ are mapped into $G$-orbits labelled by the elements in the conjugacy class of $k_2k_1^{-1}$. The number of members mapped into each $G$-orbit is equal to the order of the centralizer of $k_2k_1^{-1}$: $|Z(k_2k_1^{-1})|$. Since $K_1\otimes K_2$ are actions that preserve $\bar R$, it implies that $| R(1, 1)\rangle$ and $|R (1, gk_2k_1^{-1}g^{-1})\rangle $ are paired with the same state $|\bar{R} (1,1)\rangle$. 
The analysis can be carried out for other $G$-orbits by the replacement  $(1,1) \to (1,r)$.  
For another pair $(k_1^{'}, k_2^{'})$ in the group $K_1 \otimes K_2$ satisfying $k_2^{'}k_1^{'-1}=k_2k_1^{-1}$,  the re-distribution of members of a $G$-orbit would be identical to that resulting from  the action of $(k_1, k_2)$.
If $k_2^{'}k_1^{'-1}$ and $k_2k_1^{-1}$ belong to different conjugacy classes, $(k_1^{'}, k_2^{'})$ will map the members in 
the $G$-orbit $(1, 1)$ into $G$-orbits labelled by elements in a different conjugacy class other than the conjugacy class of $k_2k_1^{-1}$. 

Consider also the case where $k_2^{'}k_1^{'-1} = \tilde g k_2k_1^{-1} \tilde g^{-1}$ for some $\tilde g \in G$.
From the analysis above, it implies that members of a $G$-orbit would get mapped to the same destination $G$-orbits under the actions of $(k_1', k_2')$ and $(k_1,k_2)$. 
What is of note is that while these actions share the same destination $G$-orbit, the actual collection of destination members for the two actions have no overlap. That is, if the set $K_2K_1$ -- generated by all pairs $k_2 k_1^{-1}$ for $k_i\in K_i$-- contains $N$ elements belonging to the conjugacy class $C$, the number of members in the $G$-orbit $(1, 1)$ mapped into the $G$-orbit labelled by an element of that conjugacy class would be $N\cdot |Z_C|$, where $Z_C$ denote the centralizer of a representative element in the conjugacy class $C$.  

In the following, we would like to prove this assertion. 
Suppose $\{c_i\}$ are the elements of a conjugacy class $C$ of the group $G$. Denote the centralizer of each $c_i$ by $Z(c_i)$. For every pair of $c_i$ and $c_j$, choose a specific $q_{i,j} \in G$ such that $c_i = q_{i,j} c_j q_{i,j}^{-1}$ and $q_{i,j} = q_{j,i}^{-1}$. Specifically, let $q_{i,i}=1$. Then for each $c_i$, the elements in the set $q_{j,i}Z(c_i)$ would map it to $c_j$. Therefore, we only need to show that for all the different $i$, the sets $q_{j,i}Z(c_i)$ have no overlap.

We observe that $Z(c_i) = q_{i,j}Z(c_j)q_{i,j}^{-1}$. So the set $q_{j,i}Z(c_i)$ can be written as $Z(c_j)q_{i,j}^{-1}$. For different $i$ the sets $Z(c_j)q_{i,j}^{-1}$ are the right cosets of $Z(c_j)$, so the sets $Z(c_j)q_{i,j}^{-1}$ form a partition of the group $G$, completing the proof. 

We note that there is a redundancy in $k$ when $K_2 \cap K_1$ contains more than the identity element. Collecting the observations above, we are ready to write down the reduced density matrix of region $R$.
Suppose the set $K_2K_1$ contains $N$ elements belonging to the conjugacy class $C$, then $N\cdot |Z_C|$ members of the $G$-orbit $(1, 1)$ are mapped into each $G$-orbit labelled by the elements in the conjugacy class $C$. 

The block of the reduced density matrix is
\be
\begin{aligned}
(\rho_{R})_{\text{1 block}} &= \textrm{tr}_{\bar{R}} \sum_{g,g',r, r' \in G} |\bar{R}(g, rg)\rangle|R(g,rg)\rangle \langle R(g',r'g')|\langle \bar{R}(g',r'g')|\\
&= \sum_{g,g',r,r' \in G} \langle \bar{R}(g', r'g')|\bar{R}(g,rg)\rangle |R(1,r)\rangle \langle R(1,r')|.
\end{aligned}
\ee
Thus the coefficient of the term $|R(1,r)\rangle \langle R(1,r')|$ is $\sum_{g,g' \in G} \langle \bar{R}(g', r'g')|\bar{R}(g,rg)\rangle$. From the analysis above, the summation $\sum_{g,g' \in G} \langle \bar{R}(g', r'g')|\bar{R}(g,rg)\rangle$ should equal to $N\cdot |Z_C|$ multiplied by the dimension of the set $K_1 \cap K_2$. Since every entry of the reduced density matrix has a factor of the dimension of the set $K_1 \cap K_2$, it will disappear after normalization and doesn't affect the final value of the entanglement entropy.
Therefore, the procedure of writing down the reduced density matrix can be summarised as following. We only need to count the number of elements of  $K_2K_1$ belonging to each conjugacy class of $G$. 
Then for each conjugacy class $C$ we just put the number $N\cdot |Z_C|$ on the proper places in the first line in the block of the reduced density matrix. The analysis for other $G$-orbits is similar and thus we can write down the block of the reduced density matrix line by line. Other lines are just some permutations of the first line. Recall that the full reduced density matirx is obtained by $|G|^{L-2}$ such blocks, we can calculate all of its eigenvalues within the $G^2$-orbit.

\subsubsection{Abelian groups}
For Abelian groups, the $g$ appearing in the r.h.s of (\ref{eq:conjugacy_class}) clearly cancels. 
Hence, the entire $G$-orbit $(g,g)$ is mapped to one and the same $G$-orbit labeled by $(1, k_2 k_1^{-1})$. 
This implies that the wavefunction takes the following form:
\be \label{eq:wavefn}
|\psi\rangle_{\textrm{1 block}} =  \frac{1}{|G|^2} \sum_i^{|G|/|K_1K_2|} \left(\sum^{|K_1K_2|}_p | R (1, k_p r_i) \rangle \right) \otimes \left(\sum_{s}^{|G|} | \bar R(g_s,g_s r_i) \rangle \right), 
\ee
The group element $k$ however runs over the group generated by group multiplications of the elements of $K_1$ and $K_2$. If $K_1 \subseteq K_2$ then this gives $K_2$. Otherwise it is denoted more generally by $K_1K_2$ already introduced in the discussion above.  (We note that while in a non-Abelian group $G$,  $K_1K_2$ does not generically form a group, it does in the current case of an Abelian group $G$. ). The product $k r_i$ as $k$ varies over $K$ sweeps through the right coset of $K$ in $G$. 
Thus the sum over $r_i$ only runs from $i=1 \to |G|/|K_1K_2|$, picking a representative of each distinct coset of $|K_1K_2|$.   

We thus conclude that for the case of Abelian groups, the entanglement entropy of this region reads
\be \label{eq:abelian}
S(R) = \ln (|G|^{L-2} \times |G|/|K_1K_2| )= \ln |G|^{L} - \ln (|G|  |K_1K_2|).
\ee

\subsubsection{A non-Abelian example: $G=S_3$}

To illustrate the above procedure of computing the entanglement entropy, let's consider an example with $G = S_3 = \langle x,y|x^3=1,y^2=1,xyxy=1\rangle$. Let $K_1 = \{1,y\},K_2=\{1\}$. For $k_1=y, k_2 = 1$, we have $k_2k_1^{-1} = y$. The conjugacy class of $y$ is $\{y,xy,x^2y\}$, and the order of the centralizer of $y$ is 2. So under the action of $k_1 = y$ and $k_2 = 1$, the members in the $G$-orbit $(1, 1)$ are mapped into the $G$-orbits $(1, y)$, $(1, xy)$,and $(1, x^2y)$, and the number of members mapped into each of them is $2$. So the reduced density matrix of $R$ is (up to overall normalization) that ensures $\textrm{tr} \rho_R =1$)
\be
\rho_R = 
\begin{bmatrix}
6 & 0 & 0 & 2 & 2 & 2\\
0 & 6 & 0 & 2 & 2 & 2\\
0 & 0 & 6 & 2 & 2 & 2\\
2 & 2 & 2 & 6 & 0 & 0\\
2 & 2 & 2 & 0 & 6 & 0\\
2 & 2 & 2 & 0 & 0 & 6
\end{bmatrix}.
\ee
The block repeats for $6^{L-2}$ times, and the eigenvalues of the above matrix is 12, 6, 6, 6, 6, 0. So the entanglement entropy is
\be
\begin{aligned}
S = &- 6^{L-2} \cdot (\frac{12}{12\cdot 6^{L-2}+4\cdot 6 \cdot 6^{L-2}} \ln \frac{12}{12\cdot 6^{L-2}+4\cdot 6 \cdot 6^{L-2}} \\
 &+ 4 \cdot \frac{6}{12\cdot 6^{L-2}+4\cdot 6 \cdot 6^{L-2}} \ln \frac{6}{12\cdot 6^{L-2}+4\cdot 6 \cdot 6^{L-2}}) \\
=\, &(L-1)\ln 6 - \frac{1}{3}\ln 2.
\end{aligned}
\ee
We note that for non-abelian gauge groups, we do not have a closed formula for the entanglement entropy.

\subsection{Case V: Multiple PBs}
\begin{figure}[h]
\centering
\includegraphics[width=12cm]{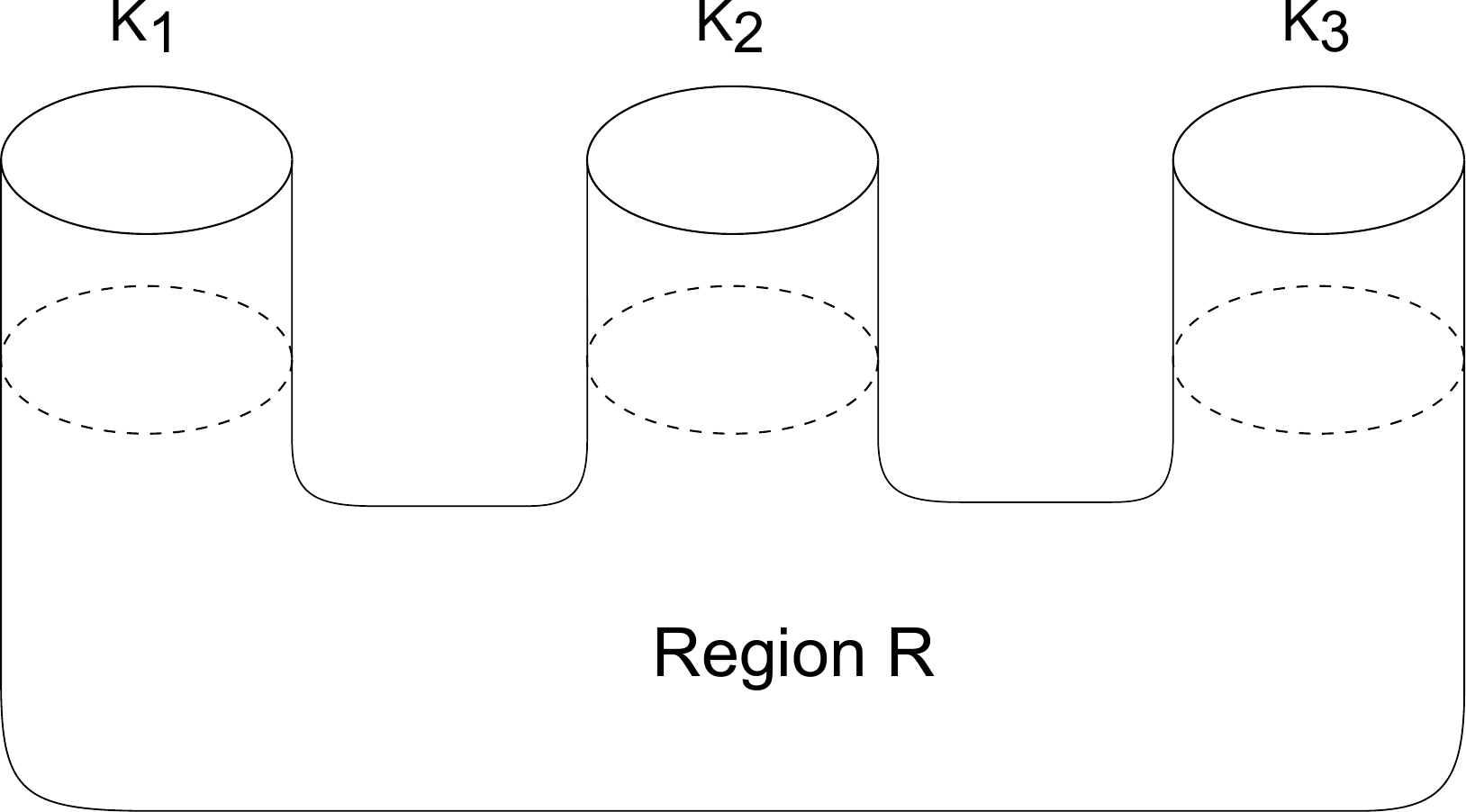}
\caption{Multiple PBs}
\label{fig:multi}
\end{figure}
If there are multiple PBs, the computation of entanglement entropy will be more complicated. Here we introduce how to compute the entanglement entropy in the case of three PBs. Suppose the PBs are characterized by subgroups $K_1, K_2, K_3\subseteq G$, and suppose the region $R$ is chosen as in Fig. \ref{fig:multi}. Thus the EB has three disconnected components We divide the EB configurations into $G^3$-orbits with each member labelled by $(g_1,g_2,g_3)$. Global shifts in $\bar{R}$ take $(g_1,g_2,g_3)$ into $(g_1k_1,g_2k_2,g_3k_3)$ and global shifts in $R$ take $(g_1,g_2,g_3)$ into $(g_1g,g_2g,g_3g)$.

First we divide a $G^3$-orbit into $|G|^2$ subsets according to the global shifts in $R$. Then each member in a subset can be labelled by $(g,r_ig,s_jg)$, where $r_i$ and $s_j$ run through the group $G$. The action of $k_1,k_2,k_3$ takes $(g,r_ig,s_jg)$ into $(gk_1,r_igk_2,s_jgk_3)$. Right multiplying the three terms by $k_1^{-1}g^{-1}$, we get $(1,r_igk_2k_1^{-1}g^{-1},s_jgk_3k_1^{-1}g^{-1})$. As $g$ runs through the group $G$, $gk_2k_1^{-1}g^{-1}$ and $gk_3k_1^{-1}g^{-1}$ run through the conjugacy class of $k_2k_1^{-1}$ and $k_3k_1^{-1}$. Due to the lack of a general relation between the centralizers of $k_2k_1^{-1}$ and $k_3k_1^{-1}$, we can't deduce a general procedure to write down the reduced density matrix. But for a specific group $G$, one can still write down the reduced density matrix and compute the entanglement entropy.

\subsection{More generic basis states with non-trivial wrapping ribbons}\label{sec:genericribbon}

In the above analysis, we have considered only the entanglement entropy of a specific reference ground state (\ref{GS}, \ref{GS1}).
On a manifold with non-contractible cycles, or open boundaries where ribbons could end, the ground state becomes degenerate. 
In the case of a cylinder, we can construct a complete basis of the ground states by wrapping magnetic ribbons around the non-contractible cycles, in addition to magnetic ribbons in the axial direction connecting the top and bottom physical boundary on the trivial state $|\Omega\rangle$ in (\ref{GS1}).
The ribbon operators are discussed in the appendix figure \ref{fig:Ribbon}. In the case of pure magnetic ribbons, we simply sum overall $g$ in the projector with equal weights. The end result is simply that we give up the projection. Since we are acting the ribbon on the reference state (\ref{GS1}), it amounts to a shift of a set of links by some group element $h$. Such a ribbon can wrap the non-contractible cycle, in which case $h\in G$. When the ribbon is one that connects the upper and lower boundaries of the cylinder, $h \in K_1\cap K_2$. We can of course have independent ribbons wrapping the non-contractible cycle and extending in the axial direction. We can thus label a reference state by a pair of group elements $(h_e, k_a)$, where $h_e$ denotes a ribbon wrapping around the non-contractible cycle, while $k_a$ extends along the axial direction.  In the presence of both, we have further restrictions of $h_e$ and $k_a$. Namely, $h_e$ and $k_a$ should commute.  

A generic ground state basis state is thus obtained by
\be \label{eq:general_state}
|\psi(h_e, k_a) \rangle=\prod_v A_v |\Omega (h_e, k_a)\rangle,
\ee
where $|\Omega(h_e, k_a) \rangle$ corresponds to a reference state obtained by action of the corresponding ribbons on the trivial reference state (\ref{GS1}). 
There is one extra complication in the presence of physical boundaries. These states are not all independent. In the presence of physical boundaries, as we have seen, a global action of $\prod{v_B} A_{v_B}(h)$ essentially generates a non-contractible ribbon corresponding to the group element $h$. Therefore, in the presence of these projectors in (\ref{eq:general_state}), we have
\be
|\psi(h_e,k_a)\rangle = |\psi(k_1 h_e k_2^{-1}, k_a)\rangle,
\ee   
for $k_i \in K_i$, where $K_1$ is the subgroup characterizing the top physical boundary, and $K_2$ the bottom physical boundary. These redundancy has already been discussed in \cite{Hung:2014tba} which analyzes the number of degenerate ground states in the presence of physical boundaries.  

To compute entanglement entropy, we note immediately that the result for each individual reference state is completely independent of the particular choice of these basis state.

A generic linear combination of these states, however, merits extra analysis. Since the situation in section \ref{sec:ringshape} is the most interesting one, we will take it as an illustration. 

To simplify the discussion further, we will restrict our explicit examples to Abelian groups. 
For given $G$ and $K_{1,2}$, we first construct the ground states. The set generated by $K_1$ and $K_2$ was denoted $K_1K_2$ in the previous section. Here, we will call it $K_1K_2= K$ which is also a subgroup of Abelian $G$. The intersection is denoted $X\equiv K_1\cap K_2$. The ground state subspace is thus $|G|/|K| \cdot |X|$ dimensions.  From the discussion above, a generic state is given by 
\be
|\Psi\rangle = \sum_{r,x} c_{r,x} |\psi( r, x)\rangle = \sum_{r,x} c_{r,x} \prod_v A_v |\Omega (r, x)\rangle,
\ee
where $r$ are group elements of the quotient group $G/K$, which can be treated as a representative of the coset of $K$. The second label $x$ denotes axial ribbon and $x\in X$.

Ribbon $x$ cuts through all the regions involved, while ribbon $r$ can lie in any of the regions.  The initial position of $r$ does not matter, since the product of projectors $A_v$ serves to deform it in all possible topologically trivial way. Therefore, without loss of generality, we can directly choose to pick the reference state with $r$ residing within region $R$. 

For each individual basis, we first analyze it exactly as we did in the previous sections, dividing each $|\Psi(r,x)\rangle $ into linear combinations of $G^2$ orbits, and organize our states into the form as in (\ref{eq:wavefn})
\be \label{eq:rxblock}
 |\psi(r,x)\rangle_{\textrm{1 block}} =  \frac{1}{|G|^2} \sum_i^{|G|/|K|} \left(\sum^{|K|}_p | R (1, k_p r_i) \rangle_{(r,x)} \right) \otimes \left(\sum_{s}^{|G|} | \bar R(g_s,g_s r_i) \rangle_{(r,x)} \right), 
\ee
where we keep track of $(r,x)$ in the subscript of the states. 
The most important ingredient in the remaining analysis is that these states $|R\,\,(\bar R)\,\,(g_1,g_2)\rangle_{(r,x)} $ may not be independent for different $(r,x)$.
There is a set of simple relations between them. There is a correspondence of states between states in a given $G^2$ orbit. Since we have used the ansatz where the ribbon $r$ resides in region $R$ in the reference state, it implies the following relation
\be \label{eq:rxredun}
| R (1, k_p r_i) \rangle_{(r_1 r_2, x)} = | R (1, k_p r_i r_1) \rangle_{(r_2, x)}, \qquad | \bar R (g, g r_i) \rangle_{(r_1 , x)} = | \bar R ( g, g r_i) \rangle_{(r_2, x)},
\ee
for all $r_i \in K$. Note that states in different $x$ sectors are immediately different and orthogonal, following from topological reasons -- that the axial ribbon $x$ always penetrates each region an even number of times i.e. whatever goes in comes out. 

Then we would like to obtain the reduced density matrix. Tracing out $\bar R$, it gives
\be
\rho_R = \sum_{x,r_a,r_b} (c_{r_a,x} c^*_{r_b,x} ) \textrm{tr}_{\bar R}  |\psi(r_a,x)\rangle \langle \psi(r_b,x) |.
\ee
As we already anticipated above, the reduced density matrix is diagonal in $x$.

It only remains to analyze each term for $x$ fixed.
Using (\ref{eq:rxblock}), we then have
\begin{align}
\begin{split}
&|\Psi\rangle_{\textrm{1 block}}\bigg\vert_{\textrm{fixed $x$}} =  \frac{|K|}{|G|}  \sum_{r, r_i}  c_{r,x} | R (1, r_i r) \rangle_{(1,x)}  \otimes |\bar R (r_i)\rangle \\
&=   \frac{|K|}{|G|}  \sum_{r_i}  \left(  \sum_r c_{r,x} | R (1, r_i r ) \rangle_{(1,x)} \right) \otimes |\bar R (r_i) \rangle
\end{split}
\end{align}

where we have used (\ref{eq:rxredun}), and simplified notation by replacing
\be
\frac{1}{|G|} \sum_{s}^{|G|} | \bar R(g_s,g_s r_i) \rangle_{(r,x)}  \equiv |\bar R (r_i)\rangle, \qquad    \frac{1}{|K|}\sum^{|K|}_p| R (1, k_p r_i ) \rangle_{(r,x)} \equiv |R(1, r_i)\rangle_{r,x}.
\ee
Here, $r_i$ used to denote a representative in the coset of $K$ in (\ref{eq:rxblock}) and it is now simply denoting a group element of the quotient group $G/K$. 
One can immediately see that the state with maximal entanglement would be the one where $c_{r_1}=1$ while all other $c_{r_j}=0$. Each of the $|G|/|K|$ state would contribute to entanglement, and we recover our previous result (\ref{eq:abelian}).
The minimally entangled state would correspond to having all $c_{r_i}$ equal. Then within the $G^2$ orbit we have a direct product state. In which case the entanglement entropy is given by
\be \label{eq:minEE}
S(R) = \ln |G|^{L-2}.
\ee

As discussed in \cite{Zhang:2011jd} this is a generic method of recovering the ``anyon line'' eigen-basis. We note that there is now non-trivial dependence on the boundary conditions for the maximally entangled state. The result gets more complicated without a clean closed form expression in the case of non-Abelian theories, although the minimally entangled shares exactly the same entanglement as in (\ref{eq:minEE}).

A complete understanding of the physical interpretation of these numbers would require new examples involving more general topological orders beyond lattice gauge theories.

\subsection{A  physical interpretation} \label{sec:interp}
In the above, we explained how to recover the Schmidt decomposition by dividing EB configurations of $A_v$ into distinct orbits or sets, and with members of each set labeled by group elements $g$. We would like to comment on the meaning of these labelling. Within the same set, different configurations are related to each other by an overall action of $A_v(g)$ for any $g\in G$ along each given connected entanglement boundary component. The action of $A_v(g)$ on a closed loop is equivalent to creating a pair of closed (magnetic) ribbons of type $g$, one inside region $R$ and the other in region $\bar R$. Therefore each member in the set really is a set of entangled state, for a single connected EB
\be \sum_{g_i} \prod_{v_{b_i}} A_{v_{b_i}} (g_i) |\Omega\rangle =\sum_{g_i \in G} |R (g_i)  \rangle \otimes |\bar R(g_i)\rangle,
\ee
where $|\Omega\rangle$ is some reference state, the subscript $i$ denotes the particular connected component among all components of the EB, and $|\Omega\rangle$ is some reference state (in this case it is the ground state).  These states $|g_i\rangle$ are to denote the creation of extra magnetic ribbon in region $R$ on top of the reference state. Now, had these $|g_i\rangle$ been orthogonal to each other, the above expression is a Schmidt decomposition. The entire analysis discussed in the current paper is about deciding which of these $|R (\bar R)\,\, (g_i)\rangle$ are in fact orthogonal to each other given that the reference state is a ground state generated by all possible linear combinations of $A_v$ covered on top of the direct product state. In the case of non-trivial topologies we have to include ground state basis states built from (magnetic) ribbons wrapping non-contractible cycles which are then buried under all possible $A_v$ showered on top. But the bottom line is that the summation of $A_v$ makes some of these $|g_i\rangle_{R (\bar R)}$ linearly dependent. 

The key piece of physics is that for all other $A_v$ acting on the interior of $R$ or $\bar R$, they necessarily create ribbons that are contractible and hence topologically trivial. Therefore
creation of $g_i$ in $R (\bar R)$ can generically be undone by $A_v$ from within the respective region if the $g_i$ ribbon (and thus the particular component of the entanglement boundary) is topologically trivial. This is the gist of the physics of the topological entanglement entropy. It is determined by the Gauss constraint which can be reformulated as having a bath of magnetic ribbons rendering these $g_i$ states linearly dependent. 

What is interesting in the analysis above, where there are PBs, is that the content of this bath of magnetic ribbons get modified. By restricting to a subgroup $K$ sitting at the PB, it is physically equivalent to allowing some ribbons to be created and destroyed at the boundary. This is of course a manifestation of anyon condensation at the boundary \cite{kitaev_kong, Hung:2014tba}. This changes the bath of ribbons in $R (\bar R)$ depending on their orientation in relation to the physical boundary, ultimately modifying the topological entanglement entropy. 
The interplay of leaking of ribbons at the boundary and the bulk bath of ribbon is clearly visible in the analysis. It appears that the interplay displays an interesting and complicated pattern when it comes to a generic non-Abelian group with multiple disconnected PBs and EBs. This issue has been pointed out before in \cite{Hung:2014tba}, where they built ground state basis in the presence of multiple boundaries characterized by different anyons condensation connected by a bulk. Then, the treatment was by no means systematic, and it was dealt with in a case-by-case manner, based on the precise mutual statistics of the condensed anyons and their fusion properties. The results in the current paper may shed new light to this problem, leading to a complete understanding.

\subsection{Comments on the twisted boundaries} \label{sec:TQD}
\begin{figure}[h]
\centering
\begin{tikzpicture}
\begin{scope}[very thick,decoration={markings,mark=at position 0.5 with {\arrow{>}}}] 
\draw[postaction={decorate}] (4,2)--(2,2);
\draw[dashed, postaction={decorate}] (4,2)--(6,2);
\draw[postaction={decorate}] (4,2)--(4,4);
\node at (1,2) {$A_v(k)$};
\node[above] at (3,2) {$a$};
\node[above] at (5,2) {$b$};
\node[right] at (4,3) {$c$};
\node at (6.5,2) {$=$};

\node at (8.2,2) {$\varphi(k,a)\varphi(k,b)^{-1}$};
\draw[postaction={decorate}] (12,2)--(10,2);
\draw[dashed, postaction={decorate}] (12,2)--(14,2);
\draw[postaction={decorate}] (12,2)--(12,4);
\node[above] at (11,2) {$ka$};
\node[above] at (13,2) {$kb$};
\node[right] at (12,3) {$kc$};
\node at (14.5,2) {$.$};
\end{scope}
\end{tikzpicture}
\caption{An illustration of the action of $A_v$ operators on the physical boundary.}
\end{figure}
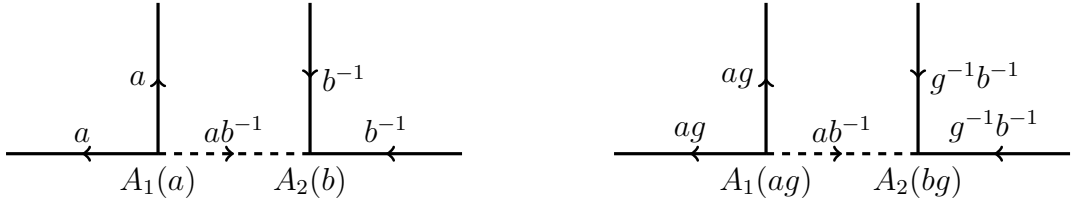
\begin{figure}
\centering
\begin{tikzpicture}
\begin{scope}[very thick,decoration={markings,mark=at position 0.5 with {\arrow{>}}}] 
\draw[postaction={decorate}] (2,0)--(0,0);
\draw[postaction={decorate}, dashed] (2,0)--(4,0);
\draw[postaction={decorate}] (2,0)--(2,2);
\draw[postaction={decorate}] (4,2)--(4,0);
\draw[postaction={decorate}] (6,0)--(4,0);
\node[above] at (1,0) {$a$};
\node[left] at (2,1) {$a$};
\node[above] at (3,0) {$ab^{-1}$};
\node[right] at (4,1) {$b^{-1}$};
\node[above] at (5,0) {$b^{-1}$};
\node[below] at (2,0) {$A_1(a)$};
\node[below] at (4,0) {$A_2(b)$};

\draw[postaction={decorate}] (10,0)--(8,0);
\draw[postaction={decorate}, dashed] (10,0)--(12,0);
\draw[postaction={decorate}] (10,0)--(10,2);
\draw[postaction={decorate}] (12,2)--(12,0);
\draw[postaction={decorate}] (14,0)--(12,0);
\node[above] at (9,0) {$ag$};
\node[left] at (10,1) {$ag$};
\node[above] at (11,0) {$ab^{-1}$};
\node[right] at (12,1) {$g^{-1}b^{-1}$};
\node[above] at (13,0) {$g^{-1}b^{-1}$};
\node[below] at (10,0) {$A_1(ag)$};
\node[below] at (12,0) {$A_2(bg)$};
\end{scope}
\end{tikzpicture}
\caption{An illustration of the action of $A_v$ located at the physical boundary before and after they are shifted by a common group element $g$. }
\label{fig:bcocycle}
\end{figure}

Next consider the case that the lattice has a physical boundary with non-trivial cocycles $\varphi:K\times K\to U(1)$. It satisfies the 2-cocycle condition:$\varphi(kl,m)\varphi(k,l)=\varphi(k,lm)\varphi(l,m)$\cite{2011CMaPh.306..663B}. The result is the same and it can be shown that the cocycles doesn't affect the computation. First consider two vertexes 1, 2. Suppose the $A_v$ operators on these two vertexes are $A_1(a)$, $A_2(b)$, originally. Then the values on the edges are $a$, $ab^{-1}$, $b^{-1}$, $c$ correspondingly, and we have a phase factor $\varphi(a,b^{-1})^{-1}$. Next we replace the $A_1(a)$, $A_2(b)$ operators by $A_1(ag)$, $A_2(bg)$ as above, where $g$ is an element of the subgroup $K$. We apply $A_2(bg)$ first and then $A_1(ag)$. This is illustrated in figure \ref{fig:bcocycle}.  The values on the edges become $ag$, $ab^{-1}$, $g^{-1}b^{-1}$, and the phase factor becomes $\varphi(ag,g^{-1}b^{-1})^{-1}$. We have
\begin{equation}
\varphi(a,g)\varphi(ag,g^{-1}b^{-1}) = \varphi(a,b^{-1})\varphi(g,g^{-1}b^{-1}).
\end{equation}
We also have
\begin{equation}
\varphi(g,g^{-1}b^{-1})\varphi(g^{-1},b^{-1}) = \varphi(1,b^{-1})\varphi(g,g^{-1}) = 1,
\end{equation}
so
\begin{equation}
\frac{\varphi(a,b^{-1})}{\varphi(ag,g^{-1}b^{-1})} = \frac{\varphi(b,g)}{\varphi(a,g)}.
\end{equation}
The new phase factor differs from the original one by a factor $\varphi(b,g)/\varphi(a,g)$. If we do the above for all the vertices on the physical boundary of region $A$, most terms will cancel and only two terms which are independent of the vertices inside region $A$ are left. So we can still use the same method to compute the entanglement entropy as if there's no non-trivial cocycle on the boundary. It is observed in \cite{Hu:2017w} that the boundary conditions are completely specified by the subgroup, and boundaries twisted by different cocycles for given subgroup can be adiabatically connected. Our results are further evidence in support of the observation in \cite{Hu:2017w}.

\section{Conclusion}

In this paper, we propose a slightly modified version of the calculation of entanglement entropy on lattice gauge theories, based on an observation locating a block diagonal structure of the reduced density matrix.
This method explicitly reproduces the known results of the entanglement entropy of lattice gauge theories on closed surfaces.

We then apply the method to compute entanglement entropy when there are non-trivial gapped PBs. We show that the entanglement entropy has very subtle dependence on the group structure.  As explained,  the physics is intimately related to the interplay of condensed anyons at different boundaries. In the case of Abelian theories, the result has a simple closed form, but not so much so for more generic non-Abelian theories. It would be interesting to explore the physical implications of these results, particularly their connection to the structure of Frobenius algebra that underlies anyon condensation, and also modular invariants that they correspond to. We note that the analysis discussed in the current paper is applicable also for the TQD. The $U(1)$ 3 cocycles would cancel in a very similar manner observed in section \ref{sec:TQD} where boundary 2-cocycles were canceled. 
It would be important to explore more general topological orders in this light beyond those describable by Dijkgraaf-Witten models. We note that some results on entanglement entropy in the presence of topological defects/interfaces have been explored in the context of Chern-Simons theories in the literature \cite{Fliss:2017wop,Gutperle:2015kmw, Wen:2016snr}.  Topological boundaries can be considered as special topological defects/interfaces. How our results are related to existing observations should be explored in greater depth in a future publication. 

Our next step would be to push these computations to higher dimensions. It is yet an open problem to have a complete classification of all possible topological boundary conditions in topological theories above 2+1 dimensions, let alone a unifying physical picture of these boundaries tantamount to the picture of anyon condensation. An understanding from the point of view of entanglement entropy should give new insights to solving the problem.

We also note that entanglement entropy can be used as a probe of higher form symmetries \cite{Hung:2018rhg}. Our results might find fresh physical interpretations in terms of a connection to anomalies -- or absence thereof -- at the physical boundary.

We will leave these interesting and exciting questions to a forth-coming publication. 

Note: We were notified of \cite{Shi:2018krj}, which appeared in January when our paper were in preparation then, after posting our paper. In \cite{Shi:2018krj}, it also explores the effect on the entanglement entropy in the presence of physical boundaries. In fact various situations considered in section 4 in the current paper have also been considered there, albeit following a different set of perspectives. In section 4.3-5 we considered the most general situations where the boundary conditions on the two (or more) boundaries of a manifold to be different. To our knowledge, this is the first instance it is considered in the literature. There are also some related results computing maximal/minimal entangled state on a cylinder in \cite{Wang:2018edf} mainly considering the $Z_2$ toric code.  We hope that our methods would supply a useful alternative to the literuature.

\appendix
\section{Open Ribbon operators and entanglement}

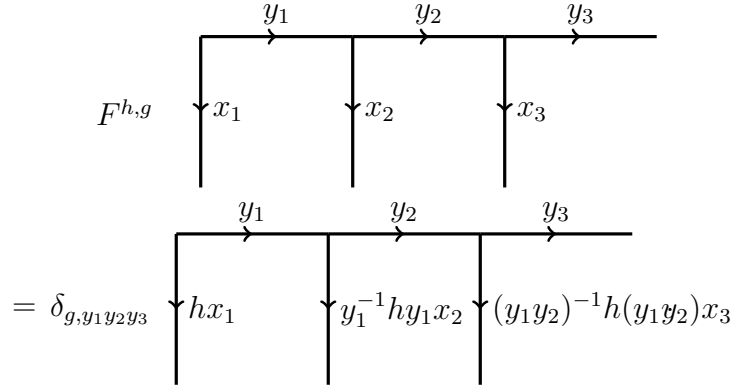
\begin{figure}[h]
\centering
\begin{tikzpicture}
\begin{scope}[very thick,decoration={markings,mark=at position 0.5 with {\arrow{>}}}] 
\draw[postaction={decorate}] (2,2)--(2,0);
\draw[postaction={decorate}] (2,2)--(4,2);
\draw[postaction={decorate}] (4,2)--(4,0);
\draw[postaction={decorate}] (4,2)--(6,2);
\draw[postaction={decorate}] (6,2)--(6,0);
\draw[postaction={decorate}] (6,2)--(8,2);
\node at (1,1) {$F^{h,g}$};
\node[right] at (2,1) {$x_1$};
\node[right] at (4,1) {$x_2$};
\node[right] at (6,1) {$x_3$};
\node[above] at (3,2) {$y_1$};
\node[above] at (5,2) {$y_2$};
\node[above] at (7,2) {$y_3$};
\end{scope}
\end{tikzpicture}

\begin{tikzpicture}
\begin{scope}[very thick,decoration={markings,mark=at position 0.5 with {\arrow{>}}}] 

\node at (7,1) {$=$};
\node at (8,1) {$\delta_{g,y_1y_2y_3}$};
\draw[postaction={decorate}] (9,2)--(9,0);
\draw[postaction={decorate}] (9,2)--(11,2);
\draw[postaction={decorate}] (11,2)--(11,0);
\draw[postaction={decorate}] (11,2)--(13,2);
\draw[postaction={decorate}] (13,2)--(13,0);
\draw[postaction={decorate}] (13,2)--(15,2);
\node[right] at (9,1) {$hx_1$};
\node[right] at (11,1) {$y_1^{-1}hy_1x_2$};
\node[right] at (13,1) {$(y_1y_2)^{-1}h(y_1y_2)x_3$};
\node[above] at (10,2) {$y_1$};
\node[above] at (12,2) {$y_2$};
\node[above] at (14,2) {$y_3$};
\node at (15.5,1) {$.$};
\end{scope}
\end{tikzpicture}
\caption{Ribbon operator.}
\label{fig:openribbon}
\end{figure}

Ribbon operators are defined in figure \ref{fig:openribbon}.

\begin{figure}[h]
\centering
\begin{tikzpicture}[scale=1.5]
\begin{scope}[very thick,decoration={markings,mark=at position 0.5 with {\arrow{>}}}]
\foreach \i in {1,...,4} {
  \draw[postaction={decorate}] (\i+1,2) -- (\i+2,2);
  \draw[postaction={decorate}] (\i+1,2) -- (\i+1,1);
}
\draw[dashed] (0,0) -- (4.2,0);
\draw[dashed] (0,0) -- (0,3);
\draw[dashed] (4.2,0) -- (4.2,3);
\draw[dashed] (0,3) -- (4.2,3);
\node[below] at (2,3) {region $R$};
\node[left] at (2,2) {$A_{v_1}(g_1)$};
\node[right] at (6,2) {$A_{v_2}(g_2)$};
\node[above] at (4.8,2.2) {$F^{h,g}$};
\end{scope}
\end{tikzpicture}
\caption{Acting the ribbon operator $F^{h,g}$ on the ground state.}
\label{fig:Ribbon}
\end{figure}
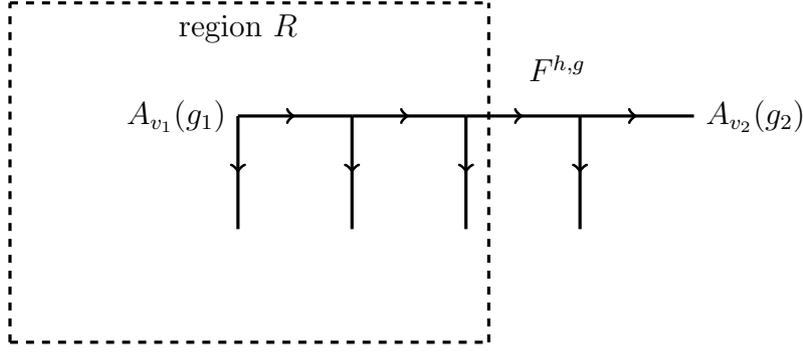

Let the ribbon operator $F^{h,g}$ act on the ground state. We choose a connected region containing one end of the ribbon as region $R$ and the complementary region as $\bar{R}$ as shown in figure \ref{fig:Ribbon}. We then compute the entanglement entropy between $R$ and $\bar{R}$.

We would like to compute the entanglement in the presence of these open ribbon operators, with one end lying in region $R$, and the other in $\bar R$. We note that this has been considered before in \cite{Delcamp:2016eya}, although the perspective and method adopted here is simpler. 
After the action of the ribbon operator $F^{h,g}$, a fixed EB configuration $\prod A_{v_i}(g_i)$ does not correspond to a direct product state. Suppose on the two ends of the ribbon the $A_v$ operators are $A_{v_1}(g_1)$ and $A_{v_2}(g_2)$. Then after the action of $F^{h,g}$ $g_1, g_2$ must satisfy $g_1g_2^{-1} = g$. So we have $g_2 = g^{-1}g_1$. We can use $g_1$ to label the state of region $R$ and use $g_2$ to label $\bar{R}$. Then for a fixed EB configuration, the state of the system can be written as
\be
|\psi\rangle = \sum_{g_1 \in G} |g_1\rangle \otimes |g^{-1}g_1\rangle.
\ee

As in the previous cases, the EB configurations can be divided into $G$-orbits. So the entanglement entropy is 
\be
S = \ln |G|^L = L \ln |G|.
\ee

Further, we can consider ribbon operators in the anyon basis. Those ribbon operators has the form\cite{Bombin}
\be
F^{RC;uv} = \sum_{n \in N_C} \bar{\Gamma}^{jj'}_R(n)F^{c_i^{-1},q_inq_{i'}^{-1}}.
\ee
Here $C$ stands for the conjugacy class, $R$ stands for the irreducible representation of a representative element $r_C$ in the conjugacy class $C$, and $N_C$ is the centralizer of $r_C$. $u$ and $v$ are two sets of integer parameters: $(i,i')$ and $(j,j')$. $\{c_i\}$ are the elements of the conjugacy class $C$. For each $c_i$ we fix a $q_i$ such that $c_i=q_ir_Cq_i^{-1}$. $\Gamma_R^{jj'}$ is the matrix element of representation $R$.

Suppose the $A_v$ operator on the end of the ribbon is $A_{v_1}(m)$. The the $A_v$ operator on the other end is $A_{v_2}(q_{i'}n^{-1}q_i^{-1}m)$. Using $m$ and $q_{i'}n^{-1}q_i^{-1}m$ to label the state of $R$ and $\bar{R}$ as previous, for a fixed EB configuration, we have
\be
|\psi\rangle = \sum_{m \in G} \sum_{n \in N_C}\Gamma^{jj'}(n^{-1}) |m\rangle\otimes|q_{i'}n^{-1}q_i^{-1}m\rangle.
\ee
We can observe that
\be
\begin{aligned}
  &\sum_{m \in q_iN_Cq_i^{-1}} \sum_{n \in N_C} \Gamma^{jj'}(n^{-1})|m\rangle \otimes |q_{i'}n^{-1}q_i^{-1}m\rangle \\
= &\sum_{m \in q_iN_Cq_i^{-1}} \sum_{n \in N_C} \Gamma^{jj'}(n^{-1}q_i^{-1}mq_i \cdot q_i^{-1}m^{-1}q_i) |m\rangle \otimes |q_{i'}n^{-1}q_i^{-1}m\rangle \\
= &\sum_{k=1}^{dim (\Gamma)} \Big(\sum_{n \in N_C} \Gamma^{jk}(n^{-1})|q_{i'}n^{-1}q_i^{-1}\rangle\quad \otimes \sum_{m \in q_iN_Cq_i^{-1}} \Gamma^{kj'}(q_i^{-1}m^{-1}q_i) |m\rangle\, \Big).
\end{aligned}
\ee
Because of the orthogonality relation in the group representation theory, the above fomulation is a Schmidt decomposition. And we can still divide the EB configurations into $G$-orbits. So the entanglement entropy is
\be
\begin{aligned}
S &= \ln \Big[ |G|^{L-1} \cdot \frac{|G|}{|N_C|}\cdot dim(\Gamma) \Big] \\
  &= L \ln |G| - \ln |G| + \ln d_i,
\end{aligned}
\ee
where $d_i=|C|\cdot dim(\Gamma)$ is the quantum dimension of the anyon type that corresponds to the ribbon operator $F^{RC;uv}$. It matches with the result in \cite{Wen:2017xwk}.

% \section{Construction of basis states using ribbon operators}

% \section{Appendix}

\section*{Acknowledgements}
We thank Arpan Bhattacharya, Yuting Hu, Zhuxi Luo, Jiaqi Lou, Ce Shen, Hongyu Wang, Gabriel Wong, Zhi Yang, Yang Zhou  for helpful discussions. We would also like to thank Juven Wang for many past conversations on the subject.
LYH acknowledges the support of Fudan University and the Thousands Young Talents Program. YW acknowledges support also from the Shanghai Pujiang Program No. KBH1512328.

\bibliographystyle{utphys}
\bibliography{bib}
\end{document}